# Wiener Reconstruction of All-Sky Spherical Harmonic Maps of the Large Scale Structure

Ofer Lahav

*Institute of Astronomy, Madingley Road, Cambridge CB3 0HA, UK*

**Abstract.** A statistical method for reconstructing large scale structure behind the Zone of Avoidance is presented. It also corrects for shot-noise and for redshift distortion in galaxy surveys. The galaxy distribution is expanded in an orthogonal set of spherical harmonics. We show that in the framework of Bayesian statistics and Gaussian random fields the $4\pi$ harmonics can be recovered and the shot-noise can be suppressed, giving the optimal picture of the underlying density field. The correction factor from observed to reconstructed harmonics turns out to be the well-known Wiener filter (the ratio of signal to signal+noise), which is also derived by requiring minimum variance. We apply the method to the 1.2 Jy IRAS survey. A reconstruction of the projected galaxy distribution confirms the connectivity of the Supergalactic Plane across the Galactic Plane (at Galactic longitude $l \sim 135°$ and $l \sim 315°$) and the Puppis cluster behind the Galactic Plane ($l \sim 240°$). The method is extended to 3-D, and is used to recover from the 1.2 Jy redshift survey the density, velocity and potential fields in the local universe.

## 1. Introduction

Where the Zone of Avoidance (ZOA) cannot be observed directly, the alternative is to reconstruct the structure in a statistical way. Previous corrections for unobserved regions in catalogues were done, somewhat ad-hoc, by populating the ZOA uniformly according to the mean density, or by interpolating the structure below and above the Galactic Plane (e.g. Lynden-Bell et al. 1989, Yahil et al. 1991, Strauss et al. 1992, Hudson 1993).

Two other problems appear in analysing the distribution of galaxies. First, if one assumes that the distribution of luminous galaxies samples an underlying smooth density field, then the discreteness of objects introduces Poisson 'shot-noise'. Second, due to peculiar velocities, redshift surveys give a distorted picture of the density field. Here we show how to recover all-sky projected density field, from galaxy surveys which suffer incomplete sky, and also how to reconstruct the density, velocity and potential fields from redshift surveys. The results presented here are based on several recent studies, by Lahav et al. (1994), Fisher et al. (1994b) and Zaroubi et al. (1994). Hoffman (this volume) discusses other aspects of the method, mainly when applied in Cartesian coordinates, and Bunn et al. (1994) recently applied a similar method to reconstruct the COBE DMR map.

The recovery of a signal from noisy and incomplete data is a classic problem of inversion, common in problems of image processing (e.g. 'seeing' or blurred HST images). A straightforward inversion is often unstable, and a regularization scheme of some sort is essential in order to interpolate where data are missing or noisy. By avoiding any prior assumptions one allows the noise and incomplete data to sometime dominate the resulting reconstruction. In the Bayesian spirit we use here raw data and a prior model to produce 'improved data'. This may raise the question to what extent a reconstruction of say the Great Attractor depends on what is assumed about the unknown nature of the spectrum of fluctuations. But the prior model does not necessarily require a speculative assumption. In the context of this work we simply require a reconstruction which obeys the constraint of the 2-point correlation function of the observed galaxy distribution, as derived from a smaller section of the sky. Using the above principles we derive a Wiener filter (the ratio of signal to signal+noise), which also follows from requiring minimum variance (e.g. Rybicki & Press 1992).

## 2. Wiener filter in theory and in practice

Let us first consider a simple pedagogical example. Assume two Gaussian variables, $x$ and $y$, with zero mean, $\langle x \rangle = \langle y \rangle = 0$ (hereafter $\langle ... \rangle$ denote ensemble average). The probability for $x$ *given* $y$ is by the rule of conditional probability

$$P(x|y) = \frac{P(x,y)}{P(y)}, \qquad (1)$$

where for Gaussian probability $P(y) \propto \exp\left(-\frac{y^2}{2<y^2>}\right)$, and the joint probability is a bivariate Gaussian $P(x,y) \propto \exp\left[-\frac{1}{2}(u^2 - 2\rho uv + v^2)/(1-\rho^2)\right]$, where $u = x/\langle x^2 \rangle^{1/2}$, $v = y/\langle y^2 \rangle^{1/2}$ and $\rho = \langle xy \rangle / \sqrt{\langle x^2 \rangle \langle y^2 \rangle}$. It then follows that the conditional probability is simply a 'shifted Gaussian'

$$P(x|y) \propto \exp\left[-\frac{1}{2}(u - \rho v)^2 / (1-\rho^2)\right]. \qquad (2)$$

The maximum *a poteriori* probability clearly occurs for $\hat{u} = \rho v$, or $\hat{x} = \frac{\langle xy \rangle}{\langle y^2 \rangle} y$. In the special case of Gaussian fields the most probable reconstruction is also the mean field (cf. Hoffman & Ribak 1991; Kaiser & Stebbins 1991). Hereafter we term them together as the 'optimal reconstruction'.

Exactly the same result for the 'optimal reconstruction' is also obtained by a different approach, by asking for the linear filter $F$ which minimizes the variance $\langle (x - Fy)^2 \rangle$. Minimizing with respect to $F$ gives indeed $\hat{F} = \frac{\langle xy \rangle}{\langle y^2 \rangle}$ and $\hat{x} = \hat{F} y$, as above. Note that although the results of the two approaches are identical, due to the quadratic nature of the functions and the linearity of the filter, the underlying assumptions are quite different. The conditional probability approach (eq. 1) requires to specify the full distribution functions (Gaussians in our case). On the other hand, the minimum variance approach only considers the second moment of the distribution function, but assumes a linear filter $F$.

Consider now the special case that $y = x + \sigma$, where $\sigma$ is a Gaussian noise uncorrelated with the true signal $x$ (hence $\langle x\sigma \rangle = 0$). It follows that the optimal estimator of the signal $\hat{x}$ given the (noisy) measurement $y$ is

$$\hat{x} = \frac{\langle x^2 \rangle}{\langle x^2 \rangle + \langle \sigma^2 \rangle} \ y \qquad (3)$$

The factor ($F$) in front of the measurement $y$ is the well-known Wiener filter commonly used in signal processing (Wiener 1949; for review see e.g. Press et al. 1992, Rybicki & Press 1992). Note that it requires *a priori* knowledge of the variances in the signal and the noise. When the noise is negligible the factor approaches unity, but when it is significant the measurement is attenuated.

A third approach is of adding a regularizing function to the usual $\chi^2$ (log-Likelihood) minimization. In fact a regularization function of the form $x^2$, motivated in our case by physical considerations of the underlying field, yields essentially a Wiener filter. Other reconstruction method use different regularization functions, e.g Maximum Entropy (e.g. Gull 1989) takes $x \ln x$.

Here we have only considered a simple example of two variables. More generally, for vectors of signal and noise and a response function $W_{\alpha\beta}$ (e.g. a 'point spread function') one can write $y_\alpha = W_{\alpha\beta}[x_\beta + \sigma_\beta]$ and derive a Wiener solution of the form $\hat{x}_\alpha = \langle x_\alpha y_\gamma \rangle \, \langle y_\gamma y_\beta^\dagger \rangle^{-1} \, y_\beta$. The Wiener formulation is greatly simplified by using orthogonal set of functions, e.g. by employing a Fourier or harmonic transforms. In particular, if $W_{\alpha\beta} = 1$ then the Wiener matrix in Fourier space is diagonal, and eq. (3) holds, but with the variables replaced by their Fourier transforms.

Figure 1 shows a 1-d example, which is also of relevance to the ZOA problem. The solid line at the bottom panel is a mock 'double-horned' HI spectrum of a galaxy (generated by H. Ferguson). To this we added real noise taken from the Dwingeloo radio-telescope, resulting in a noisy spectrum at the top panel. I then applied a Wiener filter in Fourier space, using the prior rms of the galaxy and the noise (here of course we know what they are). The dotted line in the bottom panel shows the Wiener reconstruction which indeed recovers reasonably well the galaxy spectrum. In fact, we are developing this Wiener approach as a detection algorithm for the Dwingeloo project of HI blind search behind the ZOA.

## 3. Expansion in Spherical Harmonics

Back to the large scale structure, in analysing galaxy surveys the most informative data set is of course the catalogue itself. However, it is more efficient and sometimes more insightful to compress the galaxy data. Here we use spherical harmonics to expand the galaxy distribution in a whole-sky survey. This technique has been considered for 2-D samples (e.g. Peebles 1973, Scharf et al. 1992) and more recently for analysing redshift and peculiar velocity surveys (e.g. Regös & Szalay 1989; Scharf & Lahav 1993; Lahav et al. 1993; Fisher, Scharf & Lahav 1994a; Nusser & Davis 1994; Lahav 1994 for a summary of properties).

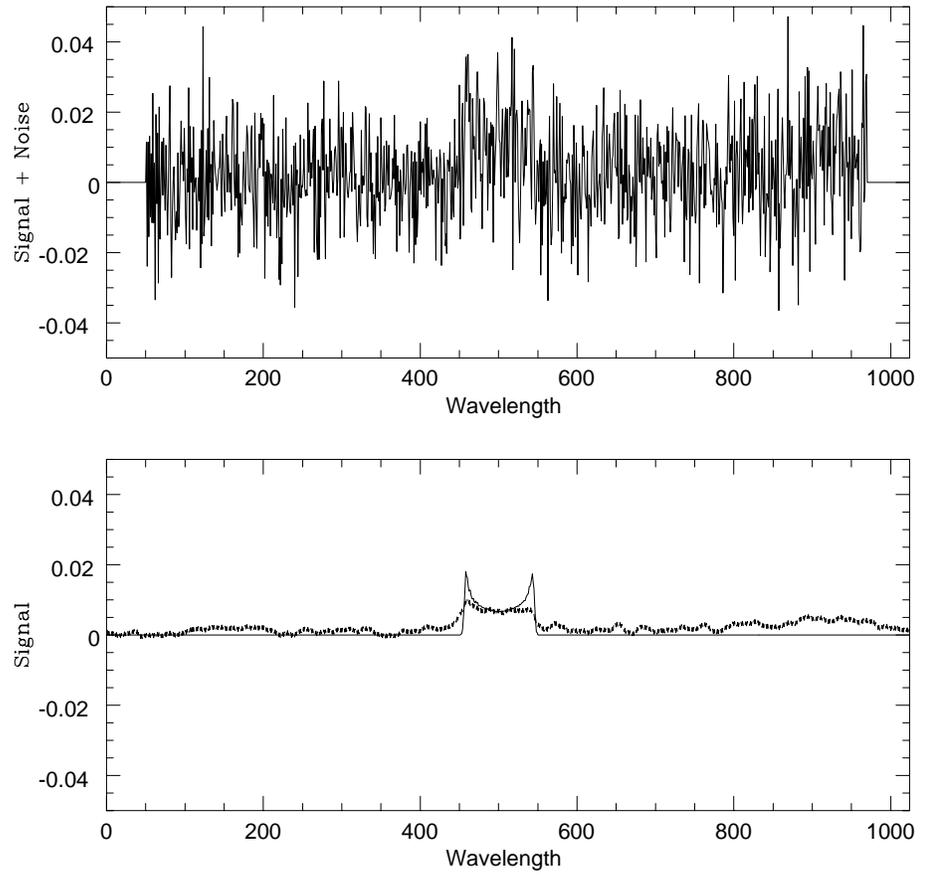

Figure 1. HI spectrum of a mock galaxy+noise (top panel) and its Wiener reconstruction (dotted line, bottom panel), compared with the original mock galaxy spectrum (solid line, bottom line). The units are arbitrary.

In projection, the density field over $4\pi$ is expanded as a sum:

$$\mathcal{S}(\theta,\phi) = \sum_l \sum_{m=-l}^{m=+l} a_{lm}\, Y_{lm}(\theta,\phi), \qquad (4)$$

where the $Y_{lm}$'s are the orthonormal set of spherical harmonics : $Y_{lm}(\theta,\phi) \propto P_l^{|m|}(\cos\theta)\,\exp(im\phi)$, where $\theta$ and $\phi$ are the spherical polar angles, and $P_l^{|m|}$'s are the associated Legendre Polynomials of degree $l$ and order $m$. The spherical harmonic analysis provides a unified language to describe the local cosmography as well as the statistical properties (e.g. the power-spectrum) of the galaxy distribution. In particular it retains both the amplitude and phase information and hence the underlying texture of the distribution.

The mean-square of harmonics can be related to the power-spectrum of mass fluctuations $P_m(k) = \langle|\delta_k|^2\rangle$ in Fourier space. In particular, the variance in harmonics measured in a flux limited redshift survey can be formulated (Fisher, Scharf & Lahav 1994a) assuming linear theory as

$$\langle|a_{lm}^S|^2\rangle = \frac{2}{\pi}\, b^2 \int dk\, k^2 P_m(k) \left|\Psi_l^R(k) + \frac{\Omega_0^{0.6}}{b}\Psi_l^C(k)\right|^2, \qquad (5)$$

where $\Psi_l^R(k)$ and $\Psi_l^C(k)$ are 'window functions'. By applying this relation to the 1.2 Jy redshift survey (of Fisher 1992, Strauss et al. 1992) we find for the combination of density and bias parameters, $\Omega_0^{0.6}/b \sim 1.0 \pm 0.3$ (assuming the observed IRAS galaxy power-spectrum).

## 4. Mask inversion using Wiener filter

We turn now to the problem of reconstructing large scale structure behind the ZOA. The problem is formulated as follows: What are the full-sky noise-free harmonics given the observed harmonics, the mask describing the unobserved region, and a prior model for the power-spectrum of fluctuations ?

The observed harmonics $c_{lm,obs}$ (with the masked regions filled in uniformly according to the mean) are related to the underlying 'true' whole-sky harmonics $a_{lm}$ by (cf. Peebles 1980, eq. 46.33)

$$c_{lm,obs} = \sum_{l'}\sum_{m'} W_{ll'}^{mm'}\, [a_{l'm'} + \sigma_a] \qquad (6)$$

where the monopole term ($l' = 0$) is excluded. We have added the shot-noise $\sigma_a$ in the 'true' number-weighted harmonics $a_{lm}$'s (not in the $c_{lm}$'s). The noise variance is estimated as $\langle\sigma_a^2\rangle = \mathcal{N}$ (the mean number of galaxies per steradian, independent $l$ in this case). The harmonic transform of the mask, $W_{ll'}^{mm'}$, introduces 'cross-talk' between the different harmonics.

By analysis similar to that given in section 2 it can be shown (Lahav et al. 1994; Zaroubi et al. 1994) that the solution of this inversion problem is

$$\hat{\mathbf{a}} = \mathbf{F}\mathbf{W}^{-1}\, \mathbf{c}_{obs}, \qquad (7)$$

where the vectors **a** and $\mathbf{c}_{obs}$ represent the sets of observed harmonics $\{a_{lm}\}$ and $\{c_{lm,obs}\}$, with the diagonal Wiener matrix

$$\mathbf{F} = diag\Big\{\frac{\langle a_l^2\rangle_{th}}{\langle a_l^2\rangle_{th} + \langle \sigma_a^2\rangle}\Big\}. \qquad (8)$$

Here $\langle a_l^2\rangle_{th}$ is the cosmic variance in the harmonics, which depends on the power-spectrum (cf. eq. 5). We emphasize again that in the special case of underlying Gaussian field the most probable field, the mean field and the minimum variance Wiener filter are all identical. The scatter in the reconstruction is at least as important, and it can also be written analytically for Gaussian random fields.

Even if the sky coverage is $4\pi$ (**W** = **I**), the Wiener filter is essential to reveal the optimal underlying 'continuous' density field, cleaned of noise. In the absence of other prior information on the location of clusters and voids, the correction factor is 'isotropic' per $l$, i.e. independent of $m$, so in the case of full sky coverage, only the amplitudes are affected by the correction, but not the relative phases. For example, the dipole direction is not affected by the shot-noise, only its amplitude. But of course, if the sky coverage is incomplete, both the amplitudes and the phases are corrected. The reconstruction also depends on number of observed and desired harmonics. Note also that the method is *non*-iterative. Since the Wiener factor is less than unity, applying it iteratively will result in zero signal !

## 5. Reconstruction of the projected IRAS 1.2 Jy galaxy distribution

Here we apply the method to the sample of IRAS galaxies brighter than 1.2 Jy which includes 5313 galaxies, and covers 88 % of the sky. This incomplete sky coverage is mainly due to the Zone of Avoidance, which we model as a 'sharp mask' at Galactic latitude $|b| < 5°$. The mean number of galaxies is $\mathcal{N} = 392$ per steradian, which sets the shot-noise, $\langle \sigma_a^2\rangle$. As our model for the cosmic scatter $\langle a_l^2\rangle_{th}$ we adopt a fit to the observed power spectrum of IRAS galaxies (Fisher et al. 1993).

Figure 2 shows the reconstruction of the projected IRAS 1.2 Jy sample. The Zone of Avoidance was left empty, and clearly it 'breaks' the possible chain of the Supergalactic Plane and other structures. Figure 3 shows our optimal reconstruction for $1 \leq l \leq 15$. Now the structure is seen to be connected across the Zone of Avoidance, in particular in the regions of Centaurus/Great Attractor ($l \sim 315°$), Hydra ($l \sim 275°$) and Perseus-Pisces ($l \sim 315°$), confirming the connectivity of the Supergalactic Plane. We also see the Puppis cluster ($l \sim 240°$) recovered behind the Galactic Plane. This cluster has been noticed in earlier harmonic expansion (Scharf et al. 1992) and other studies (Lahav et al. 1993 and references therein; Yamada in this volume). The other important feature of our reconstruction is the removal of shot noise all over the sky. This is particularly important for judging the reality of clusters and voids.

Comparison of our reconstruction with the one applied (using a $4\pi$ Wiener filter) to the IRAS sample in which the ZOA was filled in 'by hand' across the Galactic Plane (Yahil et al. 1991) shows good agreement. We have also used other prior realistic models and found that the reconstructions changed very little. As a more challenging test of the method we have also used an $N$-body simulation of standard Cold Dark matter (where the whole 'sky' true harmonics are known) and varied the size of the ZOA. We find that for mask larger than $|b| = 15°$ it is difficult to recover the unobserved structure. In this case extra-regularization is required, e.g. by truncating components in the Singular Value Decomposition (Press et al. 1992; Lahav et al. 1994). Clearly the success of the method depends on the interplay of three angular scales: the width of the mask, the desired resolution ($\pi/l_{max}$) and the physical correlation of structure.

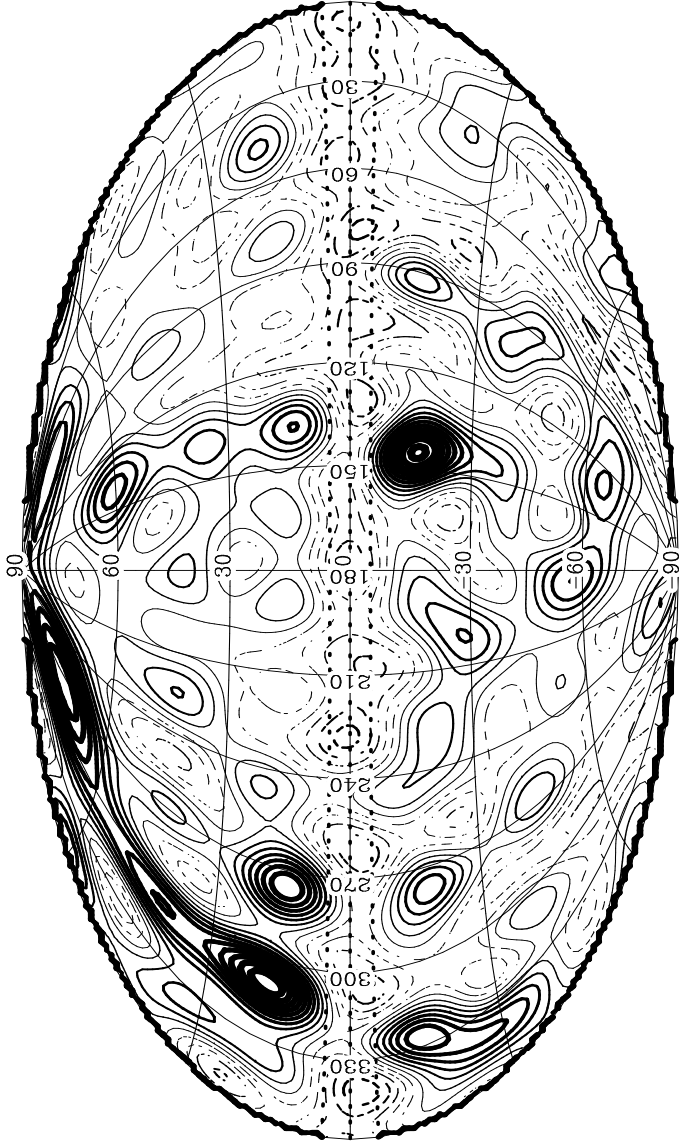

Figure 2. Harmonic expansion ($1 \leq l \leq 15$) of the projected raw IRAS 1.2 Jy data in Galactic Aitoff projection. Regions not observed, in particular $|b| < 5$ (marked by dashed lines), were left empty. The contour levels of the projected surface number density are in steps of 100 galaxies per steradian (the mean projected density is $\mathcal{N} \sim 400$ galaxies per steradian).

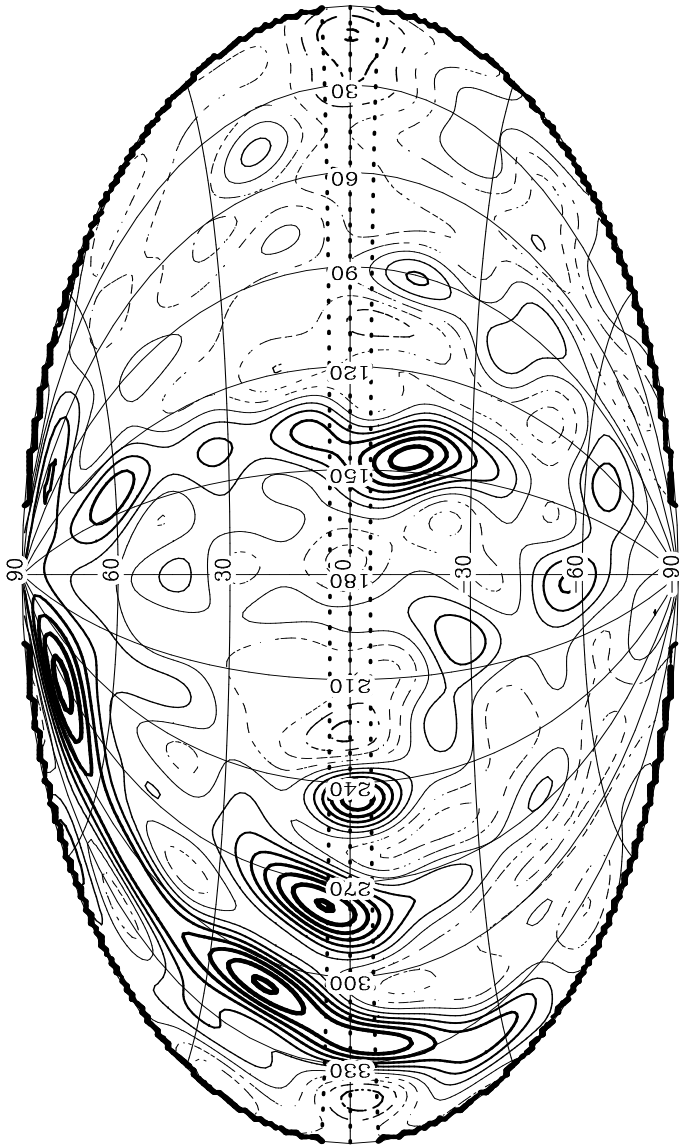

Figure 3. A whole-sky Wiener reconstruction of figure 2. The reconstruction corrects for incomplete sky coverage, as well as for the shot-noise. The reconstruction indicates that the Supergalactic Plane is connected across the Galactic Plane at Galactic longitude $l \sim 135°$ and $l \sim 315°$. The Puppis cluster stands out at the Galactic Plane at $l \sim 240°$. The horizontal dashed lines at $b = \pm 5°$ mark the major Zone of Avoidance in the IRAS sample. The contour levels are as in Figure 2. (From Lahav et al. 1994.)

## 6. 3-D reconstruction of density, velocity and potential fields

To extend the reconstruction method for analyzing redshift surveys we expand the fluctuations in the density field in spherical harmonics $Y_{lm}$ and spherical Bessel functions $j_l(z)$ (cf. Binney & Quinn 1991, Lahav 1994, Fisher et al. 1994b) :

$$\rho(\mathbf{r}) = \sum_l \sum_m \rho_{lm}(r) Y_{lm}(\hat{\mathbf{r}}) = \sum_l \sum_m \sum_n C_{nl}\, \rho_{lmn}\ j_l(k_n r)\ Y_{lm}(\hat{\mathbf{r}})\ , \quad (9)$$

The harmonics and Bessel functions are natural for this problem as they are the eigen-functions of Poisson equation, and provide a convenient framework for dynamical calculations. The $C_{nl}$'s define the normalization.

We shall assume that the data are given within a sphere of radius $R$, such that inside the sphere the desired density fluctuation is specified by $\rho_{lm}(r)$, but for $r > R$ the fluctuation is $\rho_{lm}(r) = 0$ (this simply reflects our ignorance about the density field out there; the fluctuations do not vanish of course at large distances). The Fourier $k_n$'s are chosen to ensure orthogonality of the Bessel functions, by imposing as boundary condition that the logarithmic derivative of the potential is continuous at $r = R$.

An estimator for the density field from the redshift survey is

$$\hat{\rho}^S_{lmn} = \sum_{gal} \frac{1}{\phi(s)}\, j_l(k_n s)\ Y^*_{lm}(\hat{\mathbf{r}}), \quad (10)$$

where the sum is over galaxies with $r < R$, and $\phi(r)$ is the radial selection function.

Assuming $4\pi$ coverage, two corrections are needed in order to convert the redshift space coefficients to noise-free coefficients in real space $\rho^R_{lmn}$. It is shown in detail in Fisher et al. (1994b) that this can be done by first correcting the density coefficients in redshift space for the distortion (assuming linear theory and $\Omega^{0.6}/b$; cf. eq. 5), and then applying a Wiener filter to remove the shot-noise, assuming a prior for the power-spectrum. Armed with the density coefficients one can then predict (using linear theory) the peculiar velocity field due to the mass distribution represented by galaxies inside the spherical volume. The method provides a non-parametric description of the density, velocity and potential fields, which are related by simple linear transformations. Figure 4 shows a reconstruction by this method using the IRAS 1.2 Jy redshift survey, a prior IRAS power-spectrum and $\Omega^{0.6}/b = 1$. Here we have used the IRAS survey with the ZOA filled in by the interpolation of Yahil et al. (1991), which gave similar results to our Wiener projected reconstruction described above. While the 3-D method can be extended to account for the incomplete sky coverage, it was more convenient mathematically in this case (with relatively small of ZOA and harmonics $l < 15$) to use the interpolated data and to formulate the problem for $4\pi$. It is remarkable that the two major dynamical features in the map, the Great Attractor region (at $SGX \sim -3500$ km/sec; $SGY \sim 0$ km/sec in the density map) and the Perseus-Pisces supercluster (at $SGX \sim 5000$ km/sec; $SGY \sim -2000$ km/sec) are both very near the ZOA.

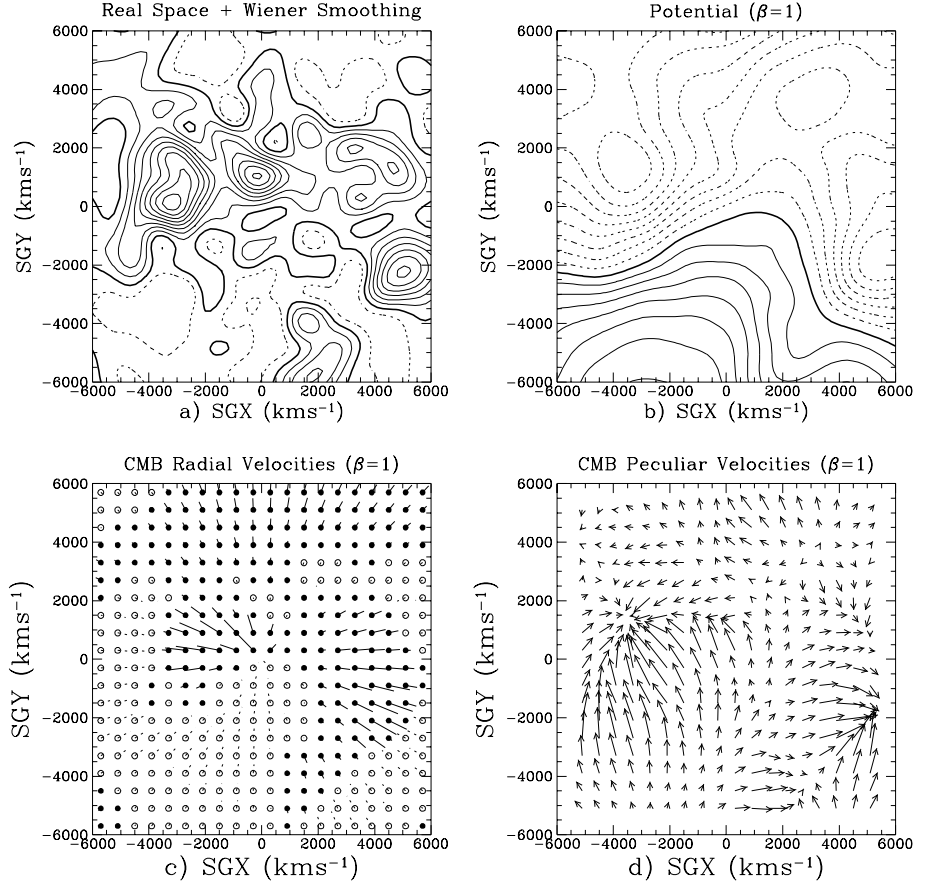

Figure 4. (a) The reconstructed real space density field from the 1.2 Jy IRAS redshift survey in the Supergalactic Plane. Contours are spaced at $\Delta\delta = 0.5$ with solid (dashed) lines denoting positive (negative) contours. The heavy solid contour corresponds to $\delta = 0$. (b) Reconstructed dimensionless gravitational potential field, $\phi(\mathbf{r})/c^2$ from the 1.2 Jy survey for $\beta = 1$. Contours are spaced at $\Delta\phi/c^2 = 5 \times 10^{-8}$. Solid (dashed) contours denote positive (negative) values with the heavy contour representing $\phi = 0$. (c) Reconstructed radial velocity field. Closed (open) dots represent positive (negative) velocities. (d) Reconstructed three dimensional peculiar velocity field. (From Fisher et al. 1994b.)

## 7. Discussion

We have presented Wiener filtering method for reconstructing the full sky density, velocity and potential fields, free of shot-noise. We have also shown that a variety of statistical approaches to the problem all lead to the same optimal Wiener estimator. The prior assumptions only depend on the observed 2-point galaxy correlation function and the nature of the shot-noise. Our assumption that the density field is Gaussian and is sampled by luminous galaxies is of course only an approximation to the real universe, but it provides a convenient framework which can be further extended. This method is to be applied to new all-sky IRAS and optical redshift surveys, and to surveys of the peculiar velocity field. The $\rho_{lmn}$'s coefficients allow objective (non-parametric) comparison of different surveys of light and mass in the local universe. As illustrated here this reconstruction technique can help in probing the ZOA, and in answering some other cosmographic questions.

**Acknowledgments.** I thank K. Fisher, Y. Hoffman, D. Lynden-Bell, C. Scharf and S. Zaroubi for their contribution to the work presented here and for many stimulating discussions.

**References**

Binney, J., & Quinn, T. 1991, MNRAS, 249, 678

Bunn, E., Fisher, K.B., Hoffman, Y., Lahav, O., Silk, J. & Zaroubi, S. 1994, preprint

Fisher, K.B. 1992, *PhD thesis*, University of California, Berkeley

Fisher, K.B., Davis, M., Strauss, M. A., Yahil, A., & Huchra, J. P. 1993, ApJ, 402, 42

Fisher, K.B., Scharf, C.A. & Lahav, O. 1994a, MNRAS, 266, 219

Fisher, K.B., Lahav, O., Hoffman, Y., Lynden-Bell, D. & Zaroubi, S. 1994b MNRAS, submitted.

Gull, S.F., 1989, in *Maximum Entropy and Bayesian Methods*, ed. J. Skilling, pg. 53 (Kluwer)

Hoffman Y. & Ribak, E. 1991, ApJ, 380, L5

Hudson, M. 1993, MNRAS, 265, 72

Kaiser, N. & Stebbins, A. 1991, in *Large Scale Structures and Peculiar Motions in the Universe*, pg. 111, eds. D.W. Latham & N. daCosta, ASP Conference Series, vol. 15, (San Francisco)

Lahav, O. 1994, in *Cosmic Velocity Fields* (Paris, July 1993), eds. F.R. Bouchet & M. Lachiéize-Rey, (Gif-sur-Tvette Cedex: Editions Frontieres), p. 205

Lahav, O., Yamada, T., Scharf, C., A. & Kraan-Korteweg, R.C. 1993, MNRAS, 262, 711

Lahav, O., Fisher, K.B., Hoffman, Y., Scharf, C.A. & Zaroubi, S. 1994, ApJ, 423, L93

Lynden-Bell, D., Lahav, O. & Burstein, D. 1989, MNRAS, 241, 325.

Nusser, A. & Davis, M. 1994, ApJ, 421, L1

Peebles, P.J.E. 1973, ApJ, 185, 413

Peebles, P.J.E. 1980, *The Large Scale Structure of the Universe*, (Princeton University Press)

Press, W. H., Teukolsky, S.A., Vetterling, W.T., & Flannery, B.P., 1992, *Numerical Recipes*, 2nd edition, (Cambridge University Press)

Scharf, C.A., Hoffman, Y., Lahav, O., & Lynden-Bell, D. 1992, MNRAS, 256, 229

Scharf, C.A., & Lahav, O. 1993, MNRAS, 264, 439

Strauss, M.A., Yahil, A., Davis, M., Huchra, J.P., & Fisher, K. 1992, ApJ, 397, 395.

Regős, E. & Szalay, A.S. 1989. ApJ, 345, 627

Rybicki, G.B., & Press, W.H. 1992, ApJ, 398, 169

Wiener, N. 1949, *Extrapolation and Smoothing of Stationary Time Series*, (New York: Wiley)

Yahil, A., Strauss, M.A., Davis, M. & Huchra, J.P., 1991, ApJ, 372, 380

Zaroubi, S., Hoffman, Y., Fisher, K.B & Lahav, O. 1994, in preparation

**Discussion**

*G. Mamon*: Can you apply spherical harmonics to one hemisphere of data ?

*O. Lahav*: This is not that practical. The Wiener reconstruction can only recover structure if the missing zone is relatively small and comparable to the correlation scale of the projected galaxy distribution. However, the nice aspect of regularization methods such as Wiener or Maximum Entropy is that they can tell you 'honestly' if the data are bad or missing, and in this case an area like the missing hemisphere is kept 'grey', i.e. at the level of the mean density.

*R. Kraan-Korteweg*: If the gap due to lack of data in whole-sky samples is too large, your reconstruction method fails. What kind of coverage of the ZOA is required to get a reliable reconstruction of the ZOA ? What kind of coverage is needed when you expand your method to 3-D and does the Wiener reconstruction allow the detection of more detailed structure compared to Potent or Hoffman's method which have very large smoothing (1200 km/sec) ?

*O. Lahav*: Our experiments with simulations, projected in the same way as the IRAS survey, indicate that reliable reconstructions are achieved only if $|b| < 15^o$. In 3-D the resolution depends on the number of Bessel radial modes, and this choice depends on the selection function, the shot-noise and the scale of non-linear redshift distortion. Practically, for current surveys 'optimal' smoothing is roughly 500 km/sec (Gaussian half-width) at distance of 4000 km/sec, but it increases with distance.

*M. Hendry*: You find that the minimum variance and maximum probability solutions are equivalent in the Gaussian case. Does this result hold for any symmetric probability distribution ? Or, perhaps a more interesting question, does the result fail for any non-symmetric distribution ?

*O. Lahav*: Indeed the Gaussian case is very special in the sense that the mean field is the same as the maximum *a posteriori* solution, and is also the same as the minimum variance solution. This equivalence breaks down for most other distribution functions, e.g. for the log-normal function.

*W. Saunders*: The Wiener filter leads to unbiased estimates of all the statistical quantities of cosmological interest - variances, velocity distortions etc. However, on a point to point basis the filtered version is always biased towards the mean by an amount dependent on the noise - surely this limits its usefulness for cosmographic purposes.

*O. Lahav*: First, The Wiener solution, when derived by the minimum variance approach, only guarantees this condition. It may, for example, bias the mean (this can be cured by subtracting the mean before the reconstruction, and adding it back at the end) and high moments. It is true that our formalism assumes that we know the noise properties in the rms, not locally. However, the formalism can be extended to accommodate other prescriptions.

13